\documentclass[prl,twocolumn,showpacs,amsmath,amssymb]{revtex4-1}
\usepackage[CJKbookmarks=true, colorlinks=true, linkcolor=blue, urlcolor=blue, citecolor=blue]{hyperref}
\usepackage{amssymb}
\usepackage{amsfonts}
\usepackage{overpic,graphicx}
\usepackage{textcomp,wasysym}
\usepackage[perpage,symbol]{footmisc}
\usepackage{lineno}
\usepackage{multirow}
\usepackage{xcolor}
\usepackage{units}
\usepackage{bigstrut}
\usepackage{array}
\usepackage{amsmath}
\usepackage[mathscr]{eucal}

\begin{document}

\title{\boldmath Search for $\bar{\Lambda}$--$\Lambda$ oscillations in the decay $J/\psi \to p K^- \bar{\Lambda}+c.c.$ }

\author{
\small
M.~Ablikim$^{1}$, M.~N.~Achasov$^{5,b}$, P.~Adlarson$^{67}$, S. ~Ahmed$^{15}$, M.~Albrecht$^{4}$, R.~Aliberti$^{28}$, A.~Amoroso$^{66A,66C}$, Q.~An$^{63,49}$, Y.~Bai$^{48}$, O.~Bakina$^{29}$
% ...
\\
\vspace{0.2cm}
(BESIII Collaboration)\\
\vspace{0.2cm} {\it
$^{1}$ Institute of High Energy Physics, Beijing 100049, People's Republic of China\\
% ...
$^{4}$ Bochum Ruhr-University, D-44780 Bochum, Germany\\
$^{5}$ Budker Institute of Nuclear Physics SB RAS (BINP), Novosibirsk 630090, Russia\\
% ...
$^{15}$ Helmholtz Institute Mainz, Staudinger Weg 18, D-55099 Mainz, Germany\\
% ...
$^{28}$ Johannes Gutenberg University of Mainz, Johann-Joachim-Becher-Weg 45, D-55099 Mainz, Germany\\
$^{29}$ Joint Institute for Nuclear Research, 141980 Dubna, Moscow region, Russia\\
% ...
$^{48}$ Southeast University, Nanjing 211100, People's Republic of China\\
$^{49}$ State Key Laboratory of Particle Detection and Electronics, Beijing 100049, Hefei 230026, People's Republic of China\\
% ...
$^{63}$ University of Science and Technology of China, Hefei 230026, People's Republic of China\\
% ...
$^{66}$ University of Turin and INFN, (A)University of Turin, I-10125, Turin, Italy; (B)University of Eastern Piedmont, I-15121, Alessandria, Italy; (C)INFN, I-10125, Turin, Italy\\
$^{67}$ Uppsala University, Box 516, SE-75120 Uppsala, Sweden\\
% ...
$^{b}$ Also at the Novosibirsk State University, Novosibirsk, 630090, Russia\\
% ...
}
\vspace{0.4cm}
}

\begin{abstract}
  We report the first search for $\bar\Lambda$--$\Lambda$ oscillations
  in the decay $J/\psi \to p K^- \bar{\Lambda} + c.c.$ by analyzing $1.31\times10^9$ $J/\psi$ events accumulated 
with the BESIII detector at the BEPCII collider.
The $J/\psi$ events are produced using
 $e^+e^-$ collisions at
  a center of mass energy $\sqrt{s}= 3.097$~GeV.  No evidence for hyperon oscillations is observed.
  The upper limit for the oscillation rate of $\bar\Lambda$ to $\Lambda$ hyperons is determined to be
  $\mathcal{P}(\Lambda)=\frac{\mathcal{B}(J/\psi\to pK^-\Lambda+c.c.)}
  {\mathcal{B}(J/\psi\to pK^-\bar\Lambda+c.c.)}<4.4\times10^{-6}$ corresponding to an oscillation parameter $\delta m_{\Lambda\bar\Lambda}$ of
  less than $3.8\times10^{-18}$~GeV at the 90\% confidence level.
\end{abstract}

\maketitle

%
%
Since the Big Bang, the Universe has evolved to a state where matter
dominates antimatter. The origin of this asymmetry remains a
mystery. In an attempt to understand this puzzle,
Sakharov~\cite{ref::sakharov} proposed three conditions that may shed
light on the asymmetry: violation of Charge ($C$) and
Charge-Parity~({\ensuremath{C\!P}}) symmetry, violation of baryon
number conservation, and deviation from thermal equilibrium.  There have been abundant experimental investigations of $C$ and {\ensuremath{C\!P}} violation with various quark decays at both non-collider and collider experiments. 
Baryon number violation (BNV) would imply the instability of the proton and, thereby, the nucleus, albeit at a time scale of the lifetime of the Universe~\cite{ref::universe}. There are many theoretical models~\cite{ref::theory} in which baryon number is not an exact symmetry of nature. For example, in some Grand Unified Theories (GUTs), the proton can decay in several ways through leptoquarks~\cite{ref::leptoquark}, such as $p\to e^+ \pi^0$.  This mechanism simultaneously breaks baryon number ($B$) and lepton number ($L$) conservation while keeping their difference $B-L$ constant. Negative results from proton decay experiments~\cite{ref::SuperK} almost rule out the entire parameter space of the simplest ($B-L$)-conserving GUTs. Therefore, it is very important to carry out an exploration of ($B-L$)-violating processes in both theory and experiment.

As reported in Ref.~\cite{ref::dutta}, recent discoveries of neutrino oscillations have made nucleon--antinucleon oscillations theoretically plausible. If small neutrino masses can be understood as a consequence of the seesaw mechanism~\cite{ref::seesaw}, it hints towards the existence of $\Delta$($B-L)=2$ interactions. There have been many experimental searches for neutron--antineutron oscillations, while few results are reported for other baryons~\cite{ref::pdg2022}.
In 2010, the authors of Ref.~\cite{ref::haibo} pointed out that the $\bar\Lambda$--$\Lambda$ oscillation phenomena can be investigated at BESIII. 
With the $\Lambda$ baryon containing a second-generation strange quark, such an investigation allows us to extend the BNV studies in proton decay and in neutron--antineutron oscillation experiments. Until now, no experimental searches for $\bar\Lambda$--$\Lambda$ oscillations have been reported.  These will be the topic of the present paper.

The time evolution of $\bar\Lambda$--$\Lambda$ oscillations is described by a Schr\"odinger-like equation (Here and elsewhere natural units $\hbar=c=1$ are used):
\begin{equation}
i\frac{\partial}{\partial t}\left(
\begin{array}{c}
 \Lambda(t)   \\
 \bar\Lambda(t)
\end{array}
\right)
=M
\left(
\begin{array}{c}
 \Lambda(t)   \\
 \bar\Lambda(t)
\end{array}
\right),
\end{equation}
where $M$ is a Hermitian matrix defined as
\begin{equation}
M=
\left(
\begin{array}{cc}
 m_{\Lambda}-\Delta E_{\Lambda}  & \delta m_{\Lambda\bar\Lambda}   \\
 \delta m_{\Lambda\bar\Lambda}  & m_{\bar\Lambda}-\Delta E_{\bar\Lambda}
\end{array}
\right),
\end{equation}
$\delta m_{\Lambda\bar\Lambda}$ is the mass splitting generated by $\Delta B=2$ transitions between $\Lambda$ and $\bar\Lambda$, $m_{\Lambda}$ ($m_{\bar\Lambda}$) is the mass of the $\Lambda$ ($\bar\Lambda$) baryon,
and $\Delta E$ is the energy splitting due to the nonzero magnetic moment of the $\Lambda$ baryon in an external magnetic field. This splitting would lead to a damping of the oscillations over time~\cite{ref::neutron-oscillation}. 
As discussed in Ref.~\cite{ref::lhcb-oscillation}, the influence of the external field is negligible if the product of the energy splitting $|\Delta E|$ and the $\Lambda$ propagation time $t$ satisfies $|\Delta E|\cdot t/2 \ll 1$. After considering the magnetic field (1.0~T) in the interaction region of the BESIII detector
and the magnetic moment of the $\Lambda$~\cite{ref::pdg2022}, the effect of the local magnetic field is estimated to be less than $0.008$, and, therefore, neglected in our case.

Starting with a beam of free $\bar\Lambda$ hyperons, the probability of generating a $\Lambda$ after time $t$, $\mathcal{P}(\Lambda, t)$, is described by 
\begin{equation}
\mathcal{P}(\Lambda, t)={\rm{sin}}^2(\delta m_{\Lambda\bar\Lambda}\cdot t)\cdot e^{-t/\tau_{\Lambda}},
\end{equation}
where the mass difference $\delta m_{\Lambda\bar\Lambda}$ is also known as the oscillation parameter, $t$ is the time at which the oscillation is observed, and $\tau_{\Lambda}=(2.632\pm0.020)\times10^{-10}\rm{~s}$~\cite{ref::pdg2022} is the lifetime of the $\Lambda$ baryon.

We measure the time-integrated oscillation rate of $\bar\Lambda\to\Lambda$ given by
\begin{equation}\label{eq::int_osc}
\mathcal{P}(\Lambda)=\frac{\int^{\infty}_{0}{\rm{sin}}^2(\delta m_{\Lambda\bar\Lambda}\cdot t)\cdot e^{-t/\tau_{\Lambda}}\cdot dt}{\int^{\infty}_{0}e^{-t/\tau_{\Lambda}}\cdot dt}.
\end{equation}
Thus, the oscillation parameter can be deduced as
\begin{equation}\label{eq::osc_par}
(\delta m_{\Lambda\bar\Lambda})^2=\frac{\mathcal{P}(\Lambda)}{2\cdot\tau_{\Lambda}^2}.
\end{equation}

%
%
This Letter reports the first search for $\bar\Lambda$--$\Lambda$ oscillations based on the decay $J/\psi\to pK^-\bar\Lambda$ (charge conjugation is implied throughout this Letter), whereby $\bar\Lambda$ baryons possibly oscillate to $\Lambda$ baryons.
This work is an experimental test of BNV with $\Delta B=2$ involving a second-generation strange quark. Furthermore, we may extract information about the oscillation parameter of the $\Lambda$ and, thereby, test the validity of related theories.

%
%
The work presented in this Letter uses $1.31\times10^9$ $J/\psi$ events that have been accumulated at a center-of-mass (c.m.) energy of $\sqrt{s}=3.097$~GeV with the BESIII detector. Details about the design and performance of the BESIII detector are given in Ref.~\cite{ref::bes3}. BESIII has a geometric acceptance covering $93$\% of the $4\pi$ solid angle. The detector consists of a helium-based multilayer drift chamber~(MDC) to track charged particles, a CsI(Tl) electromagnetic calorimeter~(EMC) to measure the energies and to reconstruct scattering angles of photons and electrons, a time-of-flight (TOF) system for charged-particle identification (PID), and a muon system for muon identification. A superconducting solenoid in an iron yoke bends the trajectories of charged particles, thereby, enabling the reconstruction of their momenta~\cite{ref::youzy}.

%
%
The analysis is performed in the framework of the BESIII Offline Software System~\cite{ref::boss} which takes care of the detector calibration, event reconstruction, data storage, and Monte Carlo (MC) simulations. Simulated data samples, generated with a {\sc geant4}-based~\cite{ref::geant4} MC package~\cite{ref::boost} including the geometric and material description of the BESIII detector, the detector response and digitization models, are used to determine the detection efficiency and to estimate the backgrounds. An inclusive MC sample of $J/\psi$ decays is generated with the {\sc kkmc}~\cite{ref::kkmc} generator at $\sqrt{s}$ = $3.097$~GeV, in which the beam energy and spread are set to the values measured experimentally at BEPCII~\cite{ref::spread}, and initial state radiation is considered. The known $J/\psi$ decays are generated with {\sc BESEvtGen}~\cite{ref::evtgen} with branching fractions (BFs) set to the world average values according to the Particle Data Group (PDG)~\cite{ref::pdg2022}, and the remaining unknown decays are modeled by Lundcharm~\cite{ref::lundcharm}.

%
%

The decay channels of interest are $J/\psi\to pK^-\bar\Lambda$ and $J/\psi\to pK^-\Lambda$, where the $\Lambda$ ($\bar\Lambda$) is reconstructed by its decay to $p\pi^-$ ($\bar p\pi^+$). We therefore select events with $p\bar p K^-\pi^+$ or $ppK^-\pi^-$ final states, respectively. We designate the events from the decay $J/\psi\to pK^-\bar\Lambda$ as {\it Right Sign} (RS) events, while the ones from $J/\psi\to pK^-\bar\Lambda\to pK^-\Lambda$ as {\it Wrong Sign} (WS) events.

%
%
All charged tracks are required to be within a polar angle ($\theta$) range of $|\cos\theta|<0.93$, where $\theta$ is defined with respect to the symmetry axis of the MDC, referred to as the $z$-axis. For charged tracks not originating from $\Lambda$ decays, the distance of closest approach to the interaction point (IP) must be less than 10\,cm along the $z$-axis ($|V_{z}|$), and less than 1\,cm in the transverse plane ($V_{xy}$). Events with exactly four selected charged tracks with zero net charge are retained for further analysis.
PID for charged tracks combines measurements of the specific energy loss d$E$/d$x$ in the MDC and from TOF to form likelihoods $\mathcal{L}(h)~(h=K,\pi, $p$)$ for each hadron $h$ hypothesis. Tracks are identified as charged kaons or protons by requiring the likelihoods for the kaon or proton hypothesis to be larger than the likelihoods of the other hypotheses. To suppress backgrounds, at least two protons and one kaon are required in each event. To optimize the detection efficiency, the pion is not explicitly identified in the analysis.
Since the $\Lambda$ baryon has a relatively long lifetime, it will travel a certain distance before it decays.  The place where the $\Lambda$ is produced is referred to as the production vertex, determined by reconstructing the electron-positron interaction point using events from Bhabha scattering. The location at which the $\Lambda$ decays into its daughter particles is referred to as the decay vertex. The flight distance between the two vertices is defined as the decay length $L$ whose average value is about 4.4~cm in this analysis. A successfully reconstructed $\Lambda$ candidate is required to pass a vertex fit in the decay of $\Lambda\to p\pi^-$ and to satisfy the requirement of $L/\sigma_L > 2$, where $\sigma_L$ is the resolution of $L$ whose average value is about 0.2~cm.  If there are more than one $p\pi^-$ combinations that meet the above criteria in one event, we keep all of them for further analysis.
To further improve the mass resolution and reduce backgrounds, 
we require all the candidate events to satisfy a four-constraint kinematic fit enforcing energy-momentum conservation for the $pK^-\bar\Lambda$ final state, \emph{i.e.} $\chi^2_{4C}<200$, where the track parameters of $\bar\Lambda$ are obtained from the $\Lambda$ vertex fit.  If more than one combination is found in an event, the one with the minimum value of $\chi^2_{4C}$ is accepted for further analysis. To further suppress mis-identification from $J/\psi$ decays with four charged tracks and large production rates, such as those with the final states $p\bar{p}K^+K^-$, $p\bar{p}\pi^+\pi^-$, $K^+K^-K^+K^-$, $\pi^+\pi^-\pi^+\pi^-$ and $K^+K^-\pi^+\pi^-$, we require that the retained candidates have the smallest $\chi^2_{4C}$ for the $pK^-\bar\Lambda$ mass assignment among the six hypotheses.

\begin{figure}[htbp]
\begin{center}
\includegraphics[width=9.5cm]{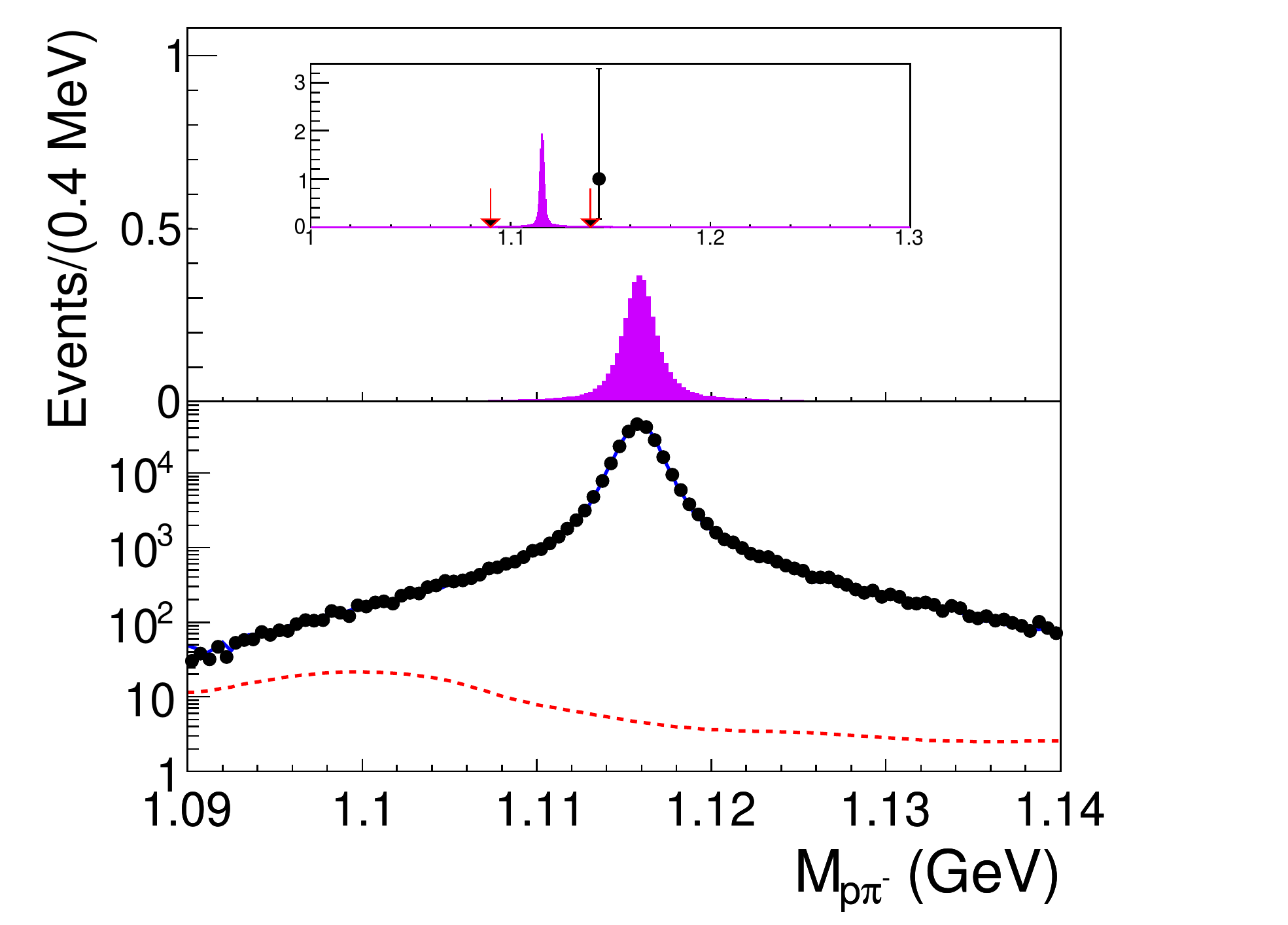}
\put(-75,170){\bf \large (a)}
\put(-75,85){\bf \large (b)}
\caption{
  Distribution of $M_{p\pi^-}$ for (a) WS events in the signal region and (inset)~over the full span, where the filled circle with error bar is from data,
  the pink filled histogram, normalized arbitrarily, stems from simulated WS signal events, and the arrows in the inset figure show the edges of the signal region;
  (b) RS events from data, where the filled circles with error bars are from data, the blue solid line represents the result of the fit and the
  dashed line shows the background contribution. 
  } \label{fig::fitting}
\end{center}
\end{figure}

%
%
The invariant mass of the proton and pion tracks, $M_{p\pi^-}$, is obtained by using the corrected energy and momentum after the kinematic fit.  Figure~\ref{fig::fitting} (b) shows the fitted $M_{p\pi^-}$ distribution. 
The background and signal responses are modeled by a non-parametric kernel estimation probability density function~\cite{Cranmer:2000du} based on the histogram from inclusive MC sample and the histogram from signal MC sample convolved with a Gaussian function to account for the difference between data and MC, respectively. 
%
The fit gives signal and background yields for RS of $N^{\rm obs}_{\rm RS}=272122\pm528$ and $N^{\rm bkg}_{\rm RS}=873\pm93$, respectively.  The signal region in $M_{p\pi^-}$ is defined as (1.09, 1.14)~GeV, which is same as the fitting range of RS events as demonstrated in Figure~\ref{fig::fitting}~(a).  No events in the WS selection survive within the signal region. Hence, we obtain $N^{\rm obs}_{\rm WS}=0$. The detection efficiencies are obtained to be $\epsilon_{\rm RS}=28.6\%$ and $\epsilon_{\rm WS}=27.8\%$ for the RS and WS processes, respectively, based on five million simulated events which are generated by the phase-space generator for $J/\psi$ decay into WS and RS final states.

%
%
The possible remaining backgrounds are studied with simulated samples and continuum data taken at an energy away from the $J/\psi$ or $\psi(3686)$ mass region.
To check the contamination from $J/\psi$ decays, we make use of an inclusive MC sample which is of the same size as the experimentally collected $J/\psi$ dataset. Only one background event is found in the WS signal region. 
To check the backgrounds from QED processes at 3.097~GeV, we analyze MC samples of the reactions $e^+e^-\to(\gamma)e^+e^-$, $e^+e^-\to(\gamma)\mu^+\mu^-$ and $e^+e^-\to q_i\bar q_j$ $(q_i=u,d,s)$ corresponding to integrated luminosities of about 0.5, 30 and 40 times of the data, respectively. No events survive the selection criteria.
Moreover, to cross-check the MC results for QED processes and to check for other potential background reactions, we investigate the data samples collected outside the vicinity of the $J/\psi$ and  $\psi(3686)$ resonances. These data samples include 
(1) $29.9$ pb$^{-1}$~\cite{ref::luminosity-scan} data taken at $\sqrt{s}=3.08$~GeV; 
(2) $44.5$ pb$^{-1}$~\cite{ref::luminosity-3773-3650} data taken at $\sqrt{s}=3.650$~GeV; 
(3) $2931.8$ pb$^{-1}$~\cite{ref::luminosity-3773-3650} data taken at $\sqrt{s}=3.773$~GeV; 
and 
(4) $1003.6$ pb$^{-1}$~\cite{ref::luminosity-scan} R scan data taken in the $\sqrt{s}$ range from 2.232 to 4.590~GeV 
excluding the energy points in the vicinity of the $J/\psi$ mass.
One background event from data taken at 3.773~GeV is found and, after normalization, leads to an expectation of $0.2$ events in the $J/\psi$ sample. For the normalization, we consider the differences in the integrated luminosities, cross sections, momenta of the 
particles, and the center-of-mass energies~\cite{ref::PRL109-042003}. Thus, a total of $1.2$ expected background events are retained due to mis-PID in the analysis of the WS process.

%
%
Since no signal event for the WS decay is observed while $1.2$ background events were estimated, we set an upper limit by utilizing a frequentist method \cite{ref::TROLKE} with unbounded profile likelihood treatment of systematic uncertainties. To determine the corresponding upper limit at the 90\% confidence level (CL) on the signal events of the WS decay $s^{90\%}_{\rm WS}$, we consider the number of signal and background events, the efficiency ($\epsilon_{\rm WS}$) and the systematic uncertainty which will be introduced later in the text. In the estimation, the number of signal and background events is assumed to follow a Poisson distribution, the efficiency is assumed to follow a Gaussian distribution, and the systematic uncertainty is incorporated as the standard deviation of the efficiency. The upper limit is calculated to be $4.2$ events at the 90\% CL.

With the assumption of no {\ensuremath{C\!P}} violation in $\Lambda$ decay in the process $J/\psi\to p K^-\bar\Lambda$, the $\bar\Lambda$--$\Lambda$ oscillation ratio which is defined in Eq.~(\ref{eq::int_osc}) can be determined by
\begin{equation}\label{eq::ratio}
\mathcal{P}(\Lambda)=\frac{ \mathcal{B}(J/\psi\to pK^-\Lambda)}
                 {\mathcal{B}(J/\psi\to pK^-\bar\Lambda) }
           =\frac{ N^{\rm obs}_{\rm WS}/\epsilon_{\rm WS}} 
                 { N^{\rm obs}_{\rm RS}/\epsilon_{\rm RS}},
\end{equation}
where $\mathcal{B}(J/\psi\to pK^-\Lambda)$ is shorthand for $\mathcal{B}(J/\psi\to pK^-\bar\Lambda 
\stackrel{oscillating}{\longrightarrow} pK^-\Lambda)$, representing the BF for 
WS channel, and $\mathcal{B}(J/\psi\to pK^-\bar\Lambda)$ is for the RS channel.
The upper limit on the $\bar\Lambda$--$\Lambda$ oscillation rate is set to be
\begin{equation}
\mathcal{P}(\Lambda)<\frac{s^{90\%}_{\rm WS}}
                 {N^{\rm obs}_{\rm RS}/\epsilon_{\rm RS}}= 4.4\times10^{-6}.
\end{equation}
As a result, the oscillation parameter $\delta m_{\Lambda\bar\Lambda}$ in Eq.~(\ref{eq::osc_par}) is calculated to be
\begin{equation}
\delta m_{\Lambda\bar\Lambda} < 3.8\times10^{-18} \rm{~GeV}.
\end{equation}

%
%
In the measurement of the ratio  $\mathcal{P}(\Lambda)$,  the systematic uncertainties associated with tracking, PID, $\Lambda$ reconstruction efficiencies, simulation of $J/\psi$ decays
and the total number of $J/\psi$ events cancel out.
The remaining systematic uncertainty originates mainly from the uncertainties in the description of the signal shape, in the efficiency determination related to the chosen fitting range and kinematical fit parameters, and in statistical uncertainties in MC samples. 
To estimate the uncertainty in the signal response, we make use of an alternative signal shape, such as a double Gaussian function or a triple Gaussian function. The difference of the signal yields (0.7\%)  between the two choices and the one used in the analysis is taken as the systematic uncertainty.
To obtain the uncertainty related to the choice of the fit range, we enlarge and shrink the range by 0.005 or 0.010~GeV, respectively, and take the relative difference of resulting BFs between the different ranges (0.6\%) as the systematic uncertainty. 
The uncertainty induced by the kinematic fit has been studied using a control sample of the $J/\psi\to\pi^+\pi^- p\bar{p}$ channel. The difference of the selection efficiencies between data and MC, with and without the fit quality requirement, is determined to be 0.2\%~\cite{ref::BAM160}.
The statistical uncertainty of the MC samples is only 0.02\% for both RS/WS processes and has been minimized by using a large number (five million) of simulated events.
The total systematic uncertainty on $\mathcal{P}(\Lambda)$ is calculated to be 1.0\% by adding all sources in quadrature. 

%
%
In summary, with $1.31\times10^9$ $J/\psi$ events collected with the BESIII detector at the BEPCII collider, the $\bar\Lambda$--$\Lambda$ oscillation process is investigated for the first time. No evidence for hyperon oscillations is observed. The upper limit on the oscillation rate is set to be $\mathcal{P}(\Lambda)<4.4\times10^{-6}$ at the $90\%$ CL.  Based on this constraint, the oscillation parameter is calculated to be $\delta m_{\Lambda\bar\Lambda}<3.8\times10^{-18}$~GeV at the $90\%$ CL corresponding to an oscillation time ($\tau_{\rm osc}=1/\delta m_{\Lambda\bar\Lambda}$) limit of $\tau_{\rm osc} > 1.7\times10^{-7}$~s at $90\%$ CL. 
Our result is comparable with the prospective constraint given in Ref.~\cite{ref::haibo}. We note that we exploited only about one-tenth of the total data sample that is currently available at BESIII. An experimental search of BNV plays a key role to understand the evolution of the Universe. In the future, at a next-generation super $\tau$-charm factory, the expected number of $J/\psi$ events can reach several trillions or larger~\cite{ref::stcf}, which can greatly improve the sensitivity on $\delta m_{\Lambda\bar\Lambda}$ to a level of at least $10^{-21}$~GeV. 
Although the upper limit on the oscillation time is much larger than the lifetime of the $\Lambda$ baryon, under special conditions, such as inside a potential well in particular hypernuclei~\cite{ref::hypernuclei}, the $\Lambda$ might exist for a much longer time to present an opportunity to obtain a better constraint.  The results presented in this Letter offer future prospects and stimulate further theoretical and experimental work.

The BESIII collaboration thanks the staff of BEPCII and the IHEP computing center for their strong support. 
This work is supported in part by National Key Basic Research Program of China under Contract No. 2015CB856700; 
National Natural Science Foundation of China (NSFC) under Contracts Nos. 12035009, 11405046, 11875170, 11475090, 11625523, 11635010, 11735014, 11822506, 11835012, 11935015, 11935016, 11935018, 11961141012; 
the Chinese Academy of Sciences (CAS) Large-Scale Scientific Facility Program; Joint Large-Scale Scientific Facility Funds of the NSFC and CAS under Contracts Nos. U1832207, U1932102, U1732263; 
CAS Key Research Program of Frontier Sciences under Contracts Nos. QYZDJ-SSW-SLH003, QYZDJ-SSW-SLH040; the CAS Center for Excellence in Particle Physics (CCEPP); 100 Talents Program of CAS; INPAC and Shanghai Key Laboratory for Particle Physics and Cosmology; ERC under Contract No. 758462; German Research Foundation DFG under Contracts Nos. 443159800, Collaborative Research Center CRC 1044, FOR 2359, FOR 2359, GRK 214; Istituto Nazionale di Fisica Nucleare, Italy; Ministry of Development of Turkey under Contract No. DPT2006K-120470; National Science and Technology fund; Olle Engkvist Foundation under Contract No. 200-0605; STFC (United Kingdom); The Knut and Alice Wallenberg Foundation (Sweden) under Contract No. 2016.0157; The Royal Society, UK under Contracts Nos. DH140054, DH160214; The Swedish Research Council; U. S. Department of Energy under Contracts Nos. DE-FG02-05ER41374, DE-SC-0012069.


\begin{thebibliography}{99}

\bibitem{ref::sakharov}  A.~D.~Sakharov,
\href{https://www.osti.gov/biblio/4449128}{JETP Lett.\  {\bf 5}, 24 (1967).}

\bibitem{ref::universe} F.~C.~Adams and G.~Laughlin, 
\href{https://doi.org/10.1103/RevModPhys.69.337}{Rev. Mod. Phys. {\bf 69}, 337 (1997).}

\bibitem{ref::theory} H.~Georgi and S.~L.~Glashow, 
\href{https://link.aps.org/doi/10.1103/PhysRevLett.32.438}{Phys. Rev. Lett. {\bf 32}, 438 (1974)}; 
P.~B.~Arnold and L.~D.~McLerran,
\href{https://link.aps.org/doi/10.1103/PhysRevD.36.581}{Phys.\ Rev.\ D {\bf 36}, 581 (1987)};
J.~A.~Harvey and M.~S.~Turner,
\href{https://link.aps.org/doi/10.1103/PhysRevD.42.3344}{Phys.\ Rev.\ D {\bf 42}, 3344 (1990)};
O.~Espinosa,
\href{https://www.sciencedirect.com/science/article/pii/055032139090473Q}{Nucl.\ Phys.\ B {\bf 343}, 310 (1990)};
S.~Dimopoulos and L.~J.~Hall,
\href{https://www.sciencedirect.com/science/article/pii/0370269388914189}{Phys.\ Lett.\ B {\bf 207}, 210 (1988)};
J.~M.~Butterworth, J.~R.~Ellis, A.~R.~Raklev and G.~P.~Salam,
\href{https://link.aps.org/doi/10.1103/PhysRevLett.103.241803}{Phys.\ Rev.\ Lett.\  {\bf 103}, 241803 (2009)}; etc.

\bibitem{ref::leptoquark} J.~C.~Pati and A.~Salam, 
\href{https://link.aps.org/doi/10.1103/PhysRevD.10.275}{Phys. Rev. D {\bf 10}, 275 (1974)}; 
\href{https://link.aps.org/doi/10.1103/PhysRevD.11.703.2}{Phys. Rev. D {\bf 11}, 703 (1975).}

\bibitem{ref::SuperK} K.~Abe {\it et al.} (Super-Kamiokande Collaboration), 
\href{https://link.aps.org/doi/10.1103/PhysRevD.95.012004}{Phys. Rev. D {\bf 95}, 012004 (2017).}

\bibitem{ref::dutta} B.~Dutta, Y.~Mimura, and R.~N.~Mohapatra, 
\href{https://link.aps.org/doi/10.1103/PhysRevLett.96.061801}{Phys. Rev. Lett. {\bf 96}, 061801 (2006).}

\bibitem{ref::seesaw} P.~Minkowski, 
\href{https://www.sciencedirect.com/science/article/pii/037026937790435X}{Phys. Lett. B {\bf 67}, 421 (1977)}; 
R.~N.~Mohapatra and G.~Senjanovic, 
\href{https://link.aps.org/doi/10.1103/PhysRevLett.44.912}{Phys. Rev. Lett. {\bf 44}, 912 (1980).}

\bibitem{ref::pdg2022} R.~L.~Workman {\it et al.} (Particle Data Group), 
\href{https://pdglive.lbl.gov/Viewer.action}{Prog. Theor. Exp. Phys. {\bf 2022}, 083C01 (2022).}

\bibitem{ref::haibo} X.-W. Kang, H.-B. Li and G.-R. Lu, 
\href{https://link.aps.org/doi/10.1103/PhysRevD.81.051901}{Phys. Rev. D {\bf 81} 051901(R) (2010).}

\bibitem{ref::neutron-oscillation} D.~G.~Phillips II {\it et al.,} 
\href{https://linkinghub.elsevier.com/retrieve/pii/S0370157315004457}{Phys. Rep. {\bf 612}, 1 (2016).}

\bibitem{ref::lhcb-oscillation} R.~Aaij {\it et al.} (LHCb Collaboration),
\href{https://journals.aps.org/prl/abstract/10.1103/PhysRevLett.119.181807}{Phys. Rev. Lett. {\bf 119}, 181807 (2017)}

\bibitem{ref::bes3} M.~Ablikim {\it et al.} (BESIII Collaboration),
\href{https://doi.org/10.1016/j.nima.2009.12.050}{Nucl. Instr. Method A {\bf 614}, 345 (2010).}

\bibitem{ref::youzy} K.~X.~Huang {\it et al.}, 
\href{https://doi.org/10.1007/s41365-022-01133-8}{Nucl. Sci. Tech. 33, 142 (2022).}

\bibitem{ref::geant4} S.~Agostinelli {\it et al.} (GEANT4 Collaboration),
\href{https://doi.org/10.1016/S0168-9002(03)01368-8}{Nucl. Instr. Method A {\bf 506}, 250 (2003).}

\bibitem{ref::boost} Z.~Y.~Deng {\it et al.}, 
\href{http://cpc.ihep.ac.cn/article/id/283d17c0-e8fa-4ad7-bfe3-92095466def1}{HEP \& NP {\bf 30}, 371 (2006).}
 
\bibitem{ref::boss} W.~D.~Li {\it et al.}, The Offline Software for the BESIII Experiment, Proceeding of CHEP2006 (Mumbai, India, 13-17 February 2006).

\bibitem{ref::kkmc} S.~Jadach, B.~F.~L.~Ward, and Z.~Was,
\href{https://journals.aps.org/prd/abstract/10.1103/PhysRevD.63.113009} {Phys. Rev. D {\bf 63}, 113009 (2001);}
\href{https://doi.org/10.1016/S0010-4655(00)00048-5}{Comput. Phys. Commun.  {\bf 130}, 260 (2000).}

\bibitem{ref::spread} J.~Y.~Zhang {\it et al.}, 
\href{http://https://doi.org/10.1016/j.nuclphysb.2018.12.023} {Nucl. Phys. B {\bf 939}, 391 (2019).}

\bibitem{ref::evtgen} D.~J.~Lange, 
\href{https://doi.org/10.1016/S0168-9002(01)00089-4} {Nucl. Instrum. Meth. A {\bf 462}, 152 (2001);}
R.~G.~Ping, 
\href{https://doi.org/10.1088/1674-1137/32/8/001}{Chin. Phys. C {\bf 32}, 599 (2008).}

\bibitem{ref::lundcharm} J.~C.~Chen, G.~S.~Huang, X.~R.~Qi, D.~H.~Zhang, and Y.~S.~Zhu, 
\href{https://journals.aps.org/prd/abstract/10.1103/PhysRevD.62.034003}{Phys. Rev. D {\bf 62}, 034003 (2000).}

\bibitem{Cranmer:2000du}
K.~S.~Cranmer,
\href{https://doi.org/doi:10.1016/S0010-4655(00)00243-5}{Comput. Phys. Commun. \textbf{136}, 198-207 (2001)}


\bibitem{ref::luminosity-scan} M.~Ablikim {\it et al.} (BESIII Collaboration), 
\href{https://doi.org/10.1088/1674-1137/41/6/063001}{Chin. Phys. C {\bf 41}, 063001 (2017).}

\bibitem{ref::luminosity-3773-3650} M.~Ablikim {\it et al.} (BESIII Collaboration), 
\href{http://hepnp.ihep.ac.cn/qikan/epaper/zhaiyao.asp?bsid=11402}{Chin. Phys. C {\bf 37}, 123001 (2013).}
\href{https://www.sciencedirect.com/science/article/pii/S0370269315008990}{Phys. Lett. B {\bf 753}, 629 (2016).}

\bibitem{ref::PRL109-042003} M.~Ablikim {\it et al.} (BESIII Collaboration), 
\href{https://journals.aps.org/prl/abstract/10.1103/PhysRevLett.109.042003}{Phys. Rev. Lett. {\bf 109}, 042003 (2012).}


\bibitem{ref::TROLKE} W.~A.~Rolke, A.~M.~Lopez and J.~Conrad, 
\href{https://doi.org/10.1016/j.nima.2005.05.068}{Nucl. Instr. Meth. A {\bf 551}, 493 (2005).} 


\bibitem{ref::BAM160} M.~Ablikim {\it et al.} (BESIII Collaboration), 
\href{https://journals.aps.org/prd/abstract/10.1103/PhysRevD.99.072006}{Phys. Rev. D {\bf 99}, 072006 (2019).}


\bibitem{ref::stcf} M.~Achasov {\it et al.} (STCF Working Group) 
\href{https://arxiv.org/pdf/2303.15790.pdf}{arXiv: 2303.15790.}

\bibitem{ref::hypernuclei} R.~H.~Dalitz and G.~Rajasekharan, 
\href{https://linkinghub.elsevier.com/retrieve/pii/0031916362904377}{Phys. Lett. {\bf 1}, 58 (1962); }
B.~I.~Abelev {\it et al.} (STAR Collaboration),  
\href{http://dx.doi.org/10.1126/science.1183980}{Science {\bf 328}, 58 (2010);}
C.~Rappold {\it et al.}, 
\href{https://doi.org/10.1016/j.physletb.2013.12.037}{Phys. Lett. B {\bf 728}, 543 (2014).}

\end{thebibliography}
\end{document}